# From the synthesis of hBN crystals to their use as nanosheets for optoelectronic devices

C. Maestre[1,2], Y. Li[1,2], V. Garnier[2], P. Steyer[2], S. Roux[3,4], A. Plaud[3,4], A. Loiseau[3], J. Barjon[4], L. Ren[5], C. Robert[5], B. Han[5], X. Marie[5], C. Journet[1*], B. Toury[1*]

*1. Laboratoire des Multimatériaux et Interfaces (LMI), UMR CNRS 5615, Univ. Lyon, Université Claude Bernard Lyon 1, F-69622 Villeurbanne, France*

*2. UCBL, CNRS, MATEIS, UMR 5510, Université de Lyon, INSA Lyon, 7 avenue Jean Capelle, Villeurbanne 69621 CEDEX, France*

*3. Laboratoire d'Etude des Microstructures, ONERA-CNRS, Université Paris-Saclay, BP 72, F-92322 Châtillon Cedex, France*

*4. Université Paris-Saclay, UVSQ, CNRS, GEMaC, F-78000 Versailles, France*

*5. Laboratoire de Physique et Chimie des Nano-Objets (LPCNO), UMR CNRS 5215, Institut National des Sciences Appliquées, 135 avenue de Rangueil, F-31077 Toulouse cedex 4, France*

[*]Authors to whom correspondence should be addressed.

**Abstract**

In the wide world of 2D materials, hexagonal boron nitride (hBN) holds a special place due to its excellent characteristics. In addition to its thermal, chemical and mechanical stability, hBN demonstrates high thermal conductivity, low compressibility, and wide band gap around 6 eV, making it promising candidate for many groundbreaking applications and more specifically for optoelectronic devices. Millimeters scale hexagonal boron nitride crystals are obtained through a disruptive dual method (PDC/PCS) consisting in a complementary coupling of the Polymer Derived Ceramics route and a Pressure-Controlled Sintering process. In addition to their excellent chemical and crystalline quality, these crystals exhibit a free exciton lifetime of 0.43 ns, as determined by time-resolved cathodoluminescence measurements, confirming their interesting optical properties. To go further in applicative fields, hBN crystals are then exfoliated, and resulting Boron Nitride NanoSheets (BNNSs) are used to encapsulate transition metal dichalcogenides (TMDs). Such van der Waals heterostructures are tested by optical spectroscopy. BNNSs do not luminesce in the emission spectral range of TMDs and the photoluminescence width of the exciton at 4K is in the range 2-3 meV. All these results demonstrate that these BNNSs are relevant for future opto-electronic applications.

**Introduction**

The continuous miniaturization of mobile devices combined with their performance and speed improvements require the development of elemental components that can meet these needs. These past years, a lot of efforts have been focused on the design of two-dimensional (2D) materials, that present specific properties related to their size reduction: electronic confinement, modifying both optical and electronic properties, high surface to volume ratio that affects the mechanical and chemical properties, reduced thickness allowing interesting flexibility, and low weight. In the wide world of 2D materials, beyond graphene, Transition Metal Dichalcogenides (TMD) and hexagonal boron nitride (hBN) hold a special place due to their excellent characteristics, especially for optoelectronic devices.[1–3] However, it is now well established that the optical, electronic and transport properties of TMD monolayers are strongly improved if the dielectric environment of the 2D layer is well controlled.[4,5,6,7,8] The best results have been obtained through encapsulation of the monolayers in high quality hBN.[9] In that sense, chemical purity and crystallinity of hBN are of prime importance. Besides, a lateral size up to hundreds of microns coupled with a thickness at the nanoscale has to be taken into account for integration into these devices. To date, most of the research groups working on the physics of 2D materials based on TMDs are still using HP/HT[10,11] high quality hBN crystals, since their exceptional quality is preserved after mechanical exfoliation.[12] However a serious bottleneck lies in their fabrication technique, which is based on severe conditions of pressure, temperature and duration, which are not easily compatible with industrial perspectives.[12,13] Besides, the use of a non-standard and heavy equipment (i.e. a massive hydraulic press) is required, which strongly limits any upscaling or industrial production. It is worth noticing that more recently, almost similar samples were achieved by an atmospheric pressure method.[13–16]

Following pioneering studies,[17–19] we have recently proposed an alternative synthesis approach conducted under quite soft conditions, involving a Polymer Derived Ceramics route and a post-Sintering process performed into a Pressure-Controlled furnace (PDC/PCS).[20] With this original coupling, large and transparent hBN single crystals up to millimeter size have been obtained through softer conditions and have led to Boron Nitride NanoSheets (BNNSs) presenting a lateral size up to few hundreds microns after exfoliation. This two-step method, which allows a BNNSs production on a larger scale and at lower cost, may represent a credible alternative to the current HP/HT source of hBN.

In this paper, we demonstrate that, whatever the considered scale, the BNNSs morphology, lattice structure and chemical composition corroborate the pure and perfect crystallization level of hBN. Further characterization by cathodoluminescence (CL) highlights the potential of these BNNSs for ideal integration into optoelectronic devices. In that sense, we show that the encapsulation of TMD monolayers with BNNSs obtained by the polymer route yield excellent optoelectronic properties, comparable to the ones achieved with the HP/HT boron nitride crystals.

**Methods**

*1)- Samples preparation*

Pure borazine is obtained from a reaction between ammonium sulfate [$(NH_4)_2SO_4$, 99% purity, Aldrich] and sodium borohydride [$NaBH_4$, 98% purity, Aldrich] in tetraglyme heated to 120 °C according to a procedure described elsewhere[21] and purified twice by distillation. The liquid state polymeric precursor is then obtained by polycondensation of borazine at 50 °C inside a pressure-sealed system under argon for 5 days, generating colorless polyborazylene (PBN) with a chemical formula $(B_{3.0}N_{3.8}H_{4.0})_n$ as given by elemental analysis. A mixture of liquid PBN and 35 wt % lithium nitride ($Li_3N$) is preceramized at 650 °C for 1 h under an inert atmosphere of $N_2$. The preceramic powder is then introduced under nitrogen into a BN crucible and thereafter transferred into a pressure-controlled furnace. The chamber is purged with three cycles ($N_2$ filling followed by pumping) to remove oxygen and moisture. The temperature and the pressure are both increased in 30 min. up to 1800 °C and up to 180 MPa with $N_2$. The dwelling time is 8 h, before cooling down to room temperature and pressure release.

hBN fragments (assembly of few single-crystalline flakes) are extracted from as-grown bulk samples by sliding with tweezers.

BNNSs are obtained by exfoliation using a polydimethylsiloxane (PDMS) dry transfer and deposited onto an oxidized silicon substrate (285 nm thick $SiO_2$).

*2)- Optical devices preparation*

The TMD based van der Waals heterostructures are fabricated by mechanical exfoliation of bulk TMD (from 2D semiconductors) and BNNs crystals using a dry-stamping technique[22]. A first layer of BNNS (bottom hBN) is mechanically exfoliated onto a freshly cleaved $SiO_2$/Si substrate. The deposition of the subsequent TMD monolayer and the second BNNS capping layer (top hBN) is obtained by repeating this procedure.

*3)- Characterization techniques*

The synthesized BN materials are characterized by optical microscopy using a Zeiss Axiophot Photomicroscope as well as by SEM using a FEG-SUPRA Zeiss 55VP microscope operating at 800V acceleration voltage. X-Ray Diffraction (XRD) is carried out with a Bruker D8 Advance diffractometer, using Cu K(α) radiation source, λ = 1.54060 Å, and the Raman spectra are recorded using a Labram HR800 spectrometer (HORIBA Jobin-Yvon) with 532 nm laser excitation wavelength. X-Ray Photoemission Spectroscopy (XPS) is performed, using a PHI Quantera SXM spectrometer with Al K(α) monochromatism radiation. Sample is kept on the holder using carbon paste. The raw XPS data are corrected using the binding energy of the C-C bond at 284.5 eV and fitted with Gaussian-Lorentzian curves. AFM images are acquired by using a CSI Nano-Observer apparatus in intermittent contact mode. The deep UV spectroscopy spectra were recorded at room temperature in a JEOL7001F field-emission-gun scanning electron microscope (SEM) coupled to a Horiba Jobin-Yvon cathodoluminescence (CL) detection system as described in Refs.[23,24] Photoluminescence (PL) and reflectance measurements were performed in a home build micro-spectroscopy set-up built around a closed-cycle, low vibration attoDry cryostat with a temperature controller (T = 4–300 K). For PL, a HeNe laser (633 nm) was used for excitation (typical excitation power of 5 µW). The white light source for reflectivity measurements is a halogen lamp with a stabilized power supply. The emitted and/or reflected light was dispersed in a spectrometer and detected by a Si-CCD camera. The excitation/detection spot diameter is ~1 µm.

**Results and Discussion**

*1)- Synthesis and characterization*

A mix of preceramized polyborazilene with a large amount of $Li_3N$ (35 w%) is introduced in a BN crucible and placed into the pressure-controlled furnace. With standard conditions applied (1800 °C, 180 MPa, 8 h), hexagonal boron nitride single crystals are formed. Actually, the use of the Li-B-N ternary system promotes the crystallization of large hBN crystals in a liquid phase.[20] XRD performed on the as-obtained samples (**figure 1**) evidences their perfect crystalline structure with all expected peaks for hBN (JCPDS 73-2095). In this case, the extreme thinness of the peaks (FWHM = $3.6 \cdot 10^{-3}$ rad) does not allow interpretation by the Debye-Scherrer method as it does not apply for crystallite sizes over 200 nm.[25]

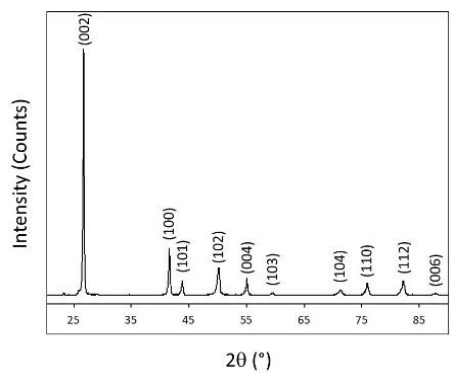

**Figure 1:** XRD pattern recorded on as-sintered hBN crystals.

These as-obtained crystals can be easily separated into independent fragments by simply using a tweezer. Their lateral size can reach the millimeter, as proven by optical (**figure 2a**) and scanning electron (**figure 2b**) images.

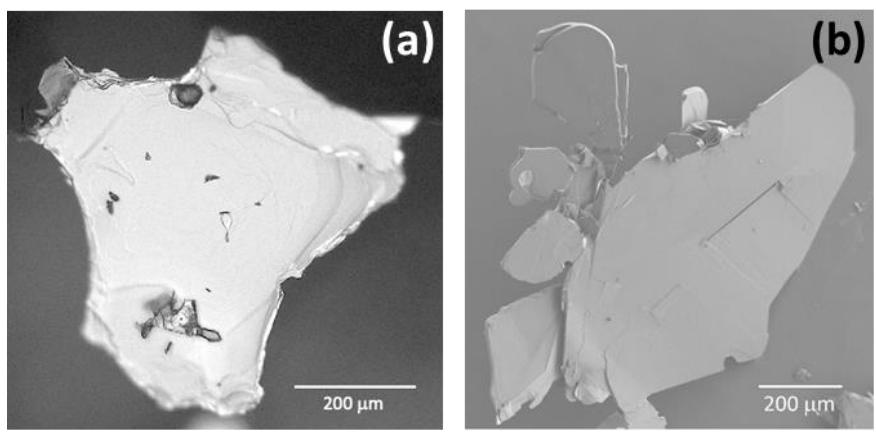

**Figure 2:** (a) Optical and (b) scanning electron microscopy images showing hBN fragments with lateral sizes up to millimeter.

The single crystal feature of these fragments is demonstrated by means of XRD (**figure 3**) that only shows hBN (002) and (004) reflections, suggesting a perfect in-plane arrangement of the hBN layers along the (002) direction.

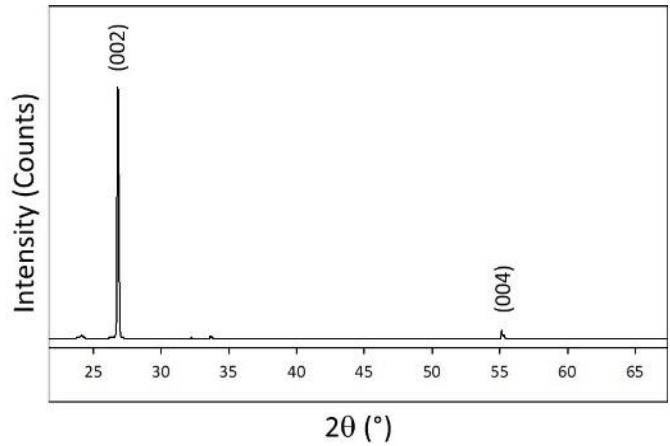

**Figure 3:** XRD pattern recorded on the h-BN single crystalline fragments on the silicon holder.

Furthermore, Raman spectroscopy measurements displayed in **Figure 4** evidence a two-phonon mode as expected for hBN crystals: an intense mode at high-frequency (1366.1 cm$^{-1}$) (**figure 4b**) which corresponds to the in-plane E$_{2g}$ optical phonon peak, and a second one in the low frequency region, at 52.6 cm$^{-1}$ (**figure 4a**) attributed to the rigid shearing oscillation between adjacent layers, which is consistent with previously reported references.[26–28] Corresponding Full Width at Half Maximum (FWHM) values of, respectively 7.6 cm$^{-1}$ and 1.3 cm$^{-1}$ are among the smallest reported in the literature for non mono-isotopic hBN[29,30] indicating, again, a hBN source of high quality.

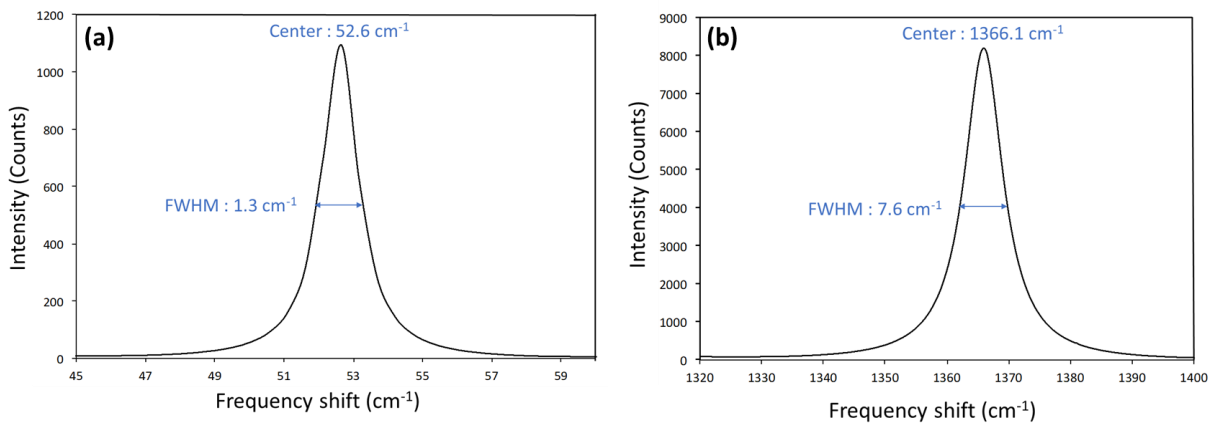

**Figure 4:** Raman spectra recorded on the hBN fragments at low (a) and high (b) frequencies.

Since Raman spectroscopy has proven its limitations for comparing high quality hBN single crystals,[31] cathodoluminescence (CL) experiments were performed in a dedicated scanning electron microscope. A spectral calibration of the detection system was used to correct the CL spectra from the low sensitivity in the UV range.[23]

**Figure 5a** shows the CL spectrum at room temperature taken from a part of a hBN crystal. The spectrum is dominated by the intrinsic radiative recombinations of hBN, i.e. free exciton recombinations with a maximum near 215 nm. A weak defect band related to carbon impurities is also detected with zero phonon line at 302 nm. The measurement of the free exciton lifetime[32] provides a meaningful benchmarking basis of the samples with two different kinds of hBN crystals: crystals grown at high temperature and high pressure (reference HP/HT crystals)[11] and crystals grown at ambient pressure and high temperature (so-called AP/HT crystals).[33,34] The free exciton lifetime was found at 0.43 ns, which indicates that the quality is lower in terms of purity and crystalline defects than for HP/HT hBN crystals (4.2 ns) but remains comparable to AP/HT ones (0.1-1.1 ns).

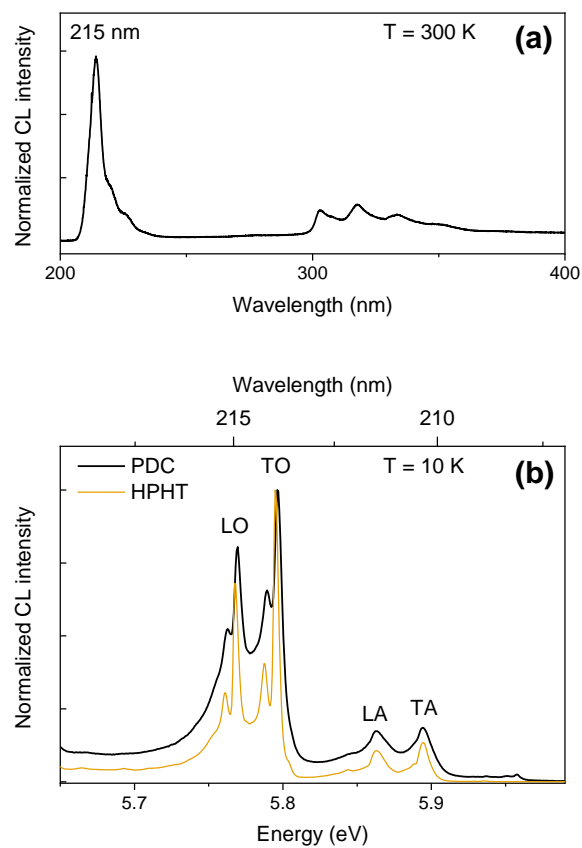

**Figure 5:** (a) CL spectrum at room temperature recorded on a hBN crystal; (b) high resolution CL spectrum at low temperature showing the free exciton recombinations assisted with different phonons. Both spectra are corrected from the spectral sensitivity of the detection system in the UV. A spectrum from a HP/HT hBN reference sample is shown for comparison.

**Figure 5b** shows a high-resolution spectrum acquired at 10K in the UV region of exciton recombinations. The phonon-assisted recombinations of the indirect exciton of hBN are clearly resolved and identified as TO, LO, TA and LA phonons. The peak linewidths are comparable to the reference hBN HP/HT crystals, plotted for comparison. These results indicate a fair quality of the hBN crystals given the soft chemical method used here.

XPS general survey recorded on a typical hBN fragment is presented in **figure 6**. The chemical composition determined with an X-ray beam of about 100 µm before and after a 2 µm $Ar^+$ etching (**inset figure 6**) indicates that surface contamination with oxygen (O1s, 532 eV) and carbon (C1s, 283 eV) is no longer present after etching, and that only expected boron (B1s, 189 eV) and nitrogen (N1s, 397 eV) species are remaining. Carbon contamination observed on the surface prior to $Ar^+$ abrasion may explain the impurity emission band visible at 300 nm on the CL spectrum.

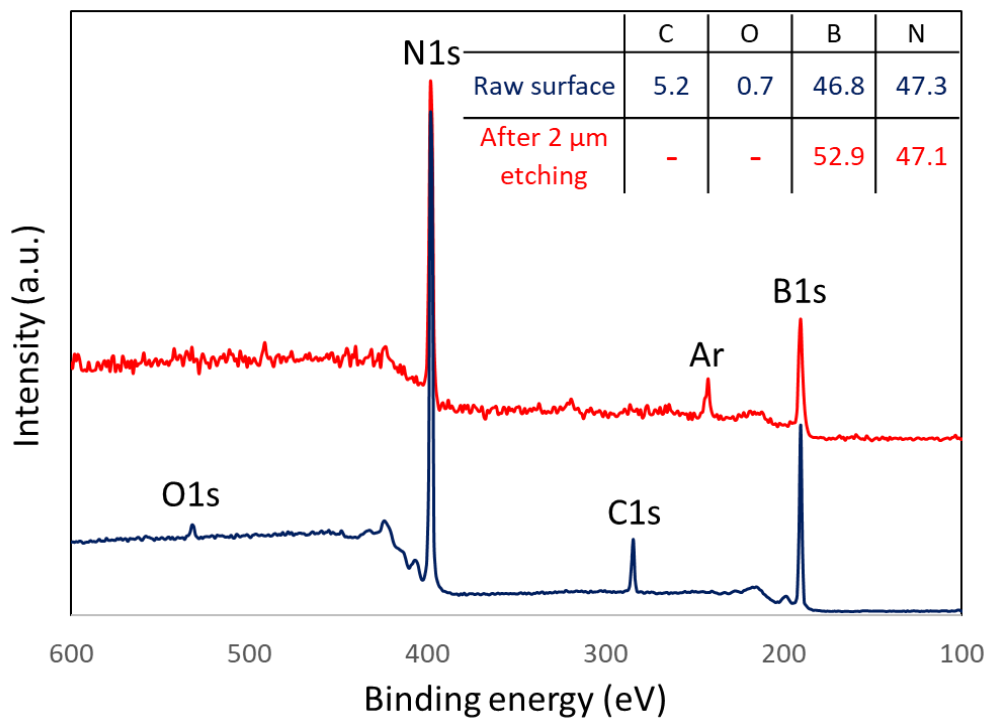

**Figure 6:** XPS general survey and elemental composition (inset) of hBN fragment before (blue) and after (red) a 2µm $Ar^+$ surface etching.

Furthermore, the B/N ratio of 0.99 measured on the raw surface proves the perfect BN stoichiometry of the crystals. As a reminder, since a preferential sputtering of nitrogen atoms occurs during the Ar$^+$ sputtering process, the B/N ratio cannot be used for quantitative composition analysis after etching. Moreover, the presence of π plasmon loss peaks at ∼9 eV away from the B1s and N1s peaks is consistent with the *sp$^2$* character of the chemical bonding within the layers.[35,36]

In summary, XRD, Raman, CL as well as XPS characterization techniques lead to the same conclusion: both the crystallinity and purity of the single crystalline fragments extracted from as-grown crystals are very much approaching the ones of HP/HT reference crystals.

For an application purpose, these fragments have then been exfoliated into BNNSs using a polydimethylsiloxane (PDMS) dry transfer and deposited onto an oxidized silicon substrate (285 nm thick SiO$_2$). SEM observations show the exfoliation efficiency since the BNNSs keep lateral dimensions of more than hundreds of microns (**figure 7a-b**) while their thickness is reduced to 2.5 - 4 nm, which corresponds to 7 to 12 stacked layers as proven by AFM (**figure 7c**).

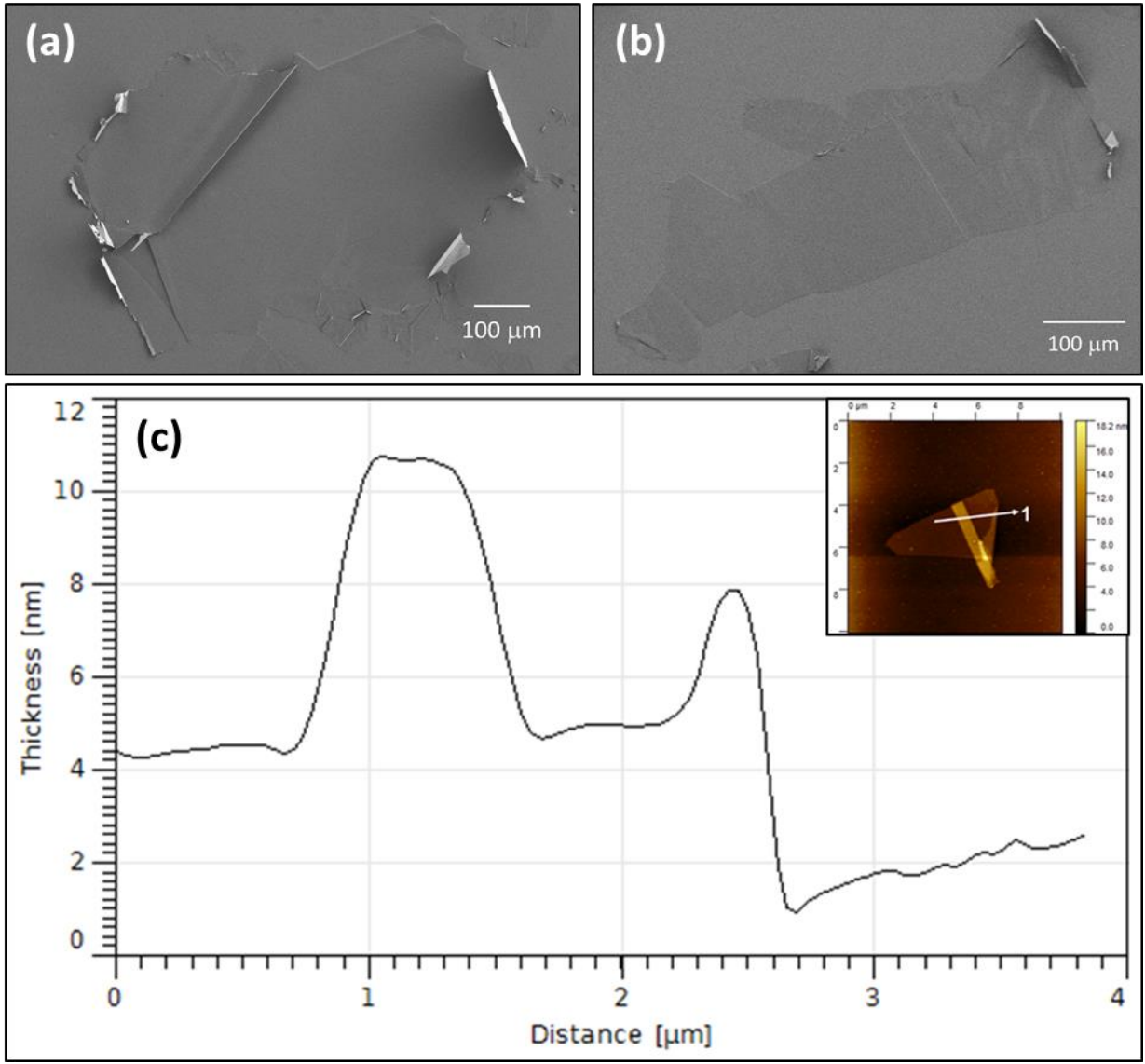

**Figure 7:** (a-b) SEM images of BNNSs of a few hundreds of microns as lateral size; (c) AFM profile obtained on BNNSs (inset).

*2)- Optoelectronic devices*

Seven devices of MoSe$_2$ and WSe$_2$ monolayers (ML) encapsulated into BNNSs have been fabricated by mechanical exfoliation and stamping. Their properties have been investigated by optical spectroscopy and they exhibit similar properties. We show the results on two of them, one based on MoSe$_2$ and the other one on WSe$_2$, in **figures 8 and 9** respectively. Figure 8a shows an optical microscopy image of the fabricated van der Waals heterostructure hBN/MoSe$_2$/hBN. Figure 8b presents the photoluminescence (PL) spectrum for a hBN/MoSe$_2$ ML/hBN structure at T=5 K. It evidences clearly two narrow peaks corresponding respectively to the recombination of neutral exciton ($X_0$) and charged exciton ($X_T$), in agreement with previous reports.[4,5] Remarkably the PL linewidth of both lines is very narrow. The FWHM of the neutral exciton line is ~ 2.7 meV, demonstrating the very high quality of the structure using BNNSs as encapsulation layers. Comparable linewidths are obtained with HP/HT hBN encapsulation[4,5,37,38]. It is worth mentioning that non-encapsulated TMD monolayers are usually characterized by PL linewidths of a few tens of meV due to the dielectric disorder around the TMD monolayer.[8,39] One can notice in figure 8b a rather large intensity of the charged exciton PL (larger than the one usually obtained with HP/HT crystals), indicating a possible doping effect of the TMD monolayer induced by the BNNS material. Figure 8c displays the PL spectrum measured at room temperature with a linewidth of about 39 meV, indicating a small inhomogeneous contribution as a consequence again of the high quality of encapsulation of the TMD monolayer with BNNSs.

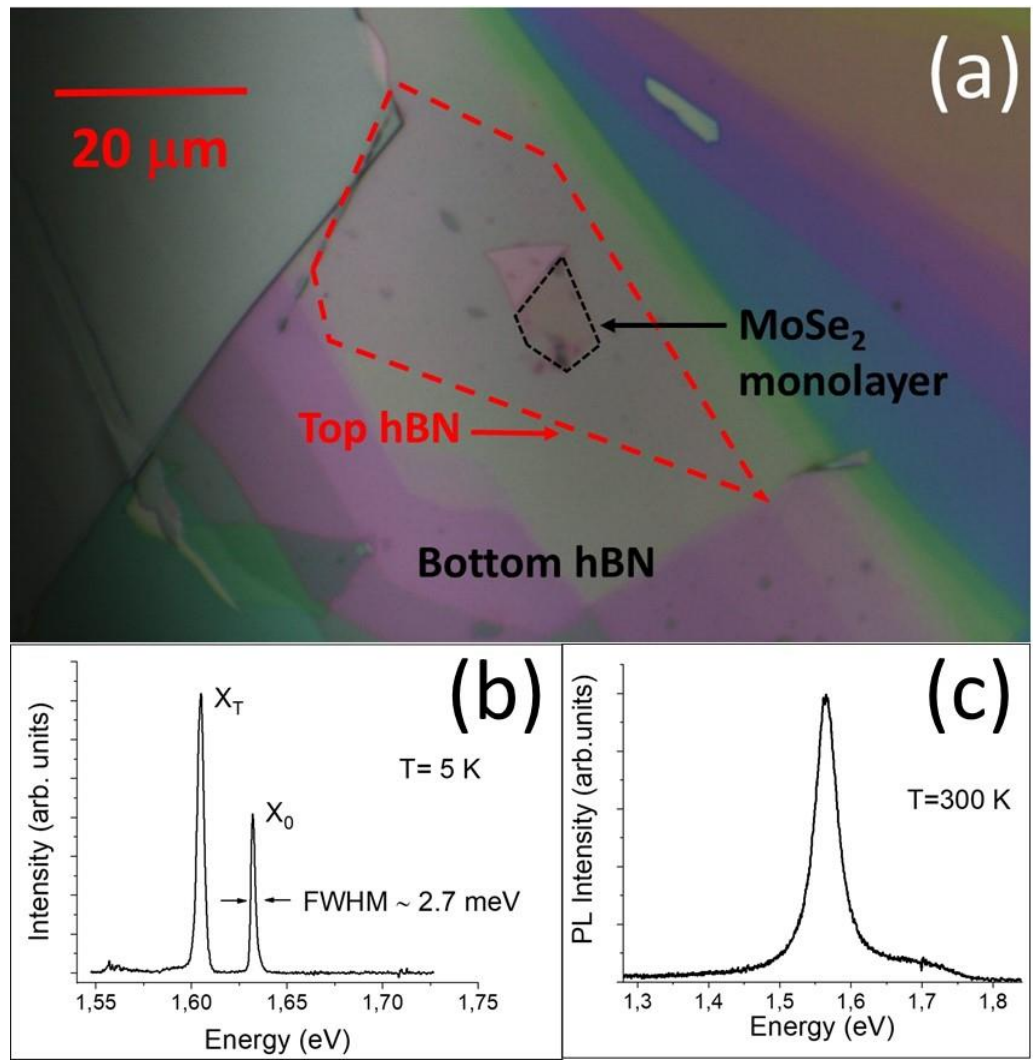

**Figure 8:** MoSe$_2$ monolayer – (a) Optical microscope image of the hBN/MoSe$_2$ monolayer/hBN heterostructure; Photoluminescence spectrum measured at (b) T=5 K and (c) T=300 K.

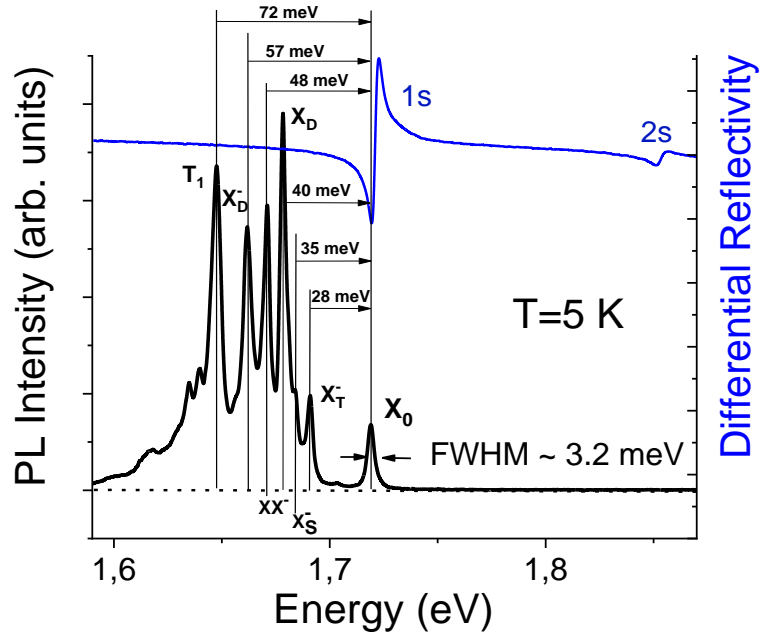

**Figure 9:** WSe$_2$ monolayer – Differential reflectivity (right axis) and Photoluminescence spectra of the hBN/WSe$_2$ monolayer/hBN heterostructure measured at T=5 K.

In order to test the BNNSs encapsulation with another TMD material, we have investigated the properties of hBN/WSe$_2$ ML/hBN structures. Figure 9 presents the differential reflectivity spectrum measured at low temperature (T=5 K). It exhibits two narrow resonances at 1.721 and 1.854 eV, corresponding to the neutral exciton $X_0$ ground state (1s) and excited state (2s) respectively. The observation of the exciton excited state attests to the high quality of the structure. Figure 9 also displays the measured PL spectrum. At high energy, we observe the neutral exciton line $X_0$, characterized by a line width of 3.2 meV (FWHM). This narrow linewidth is comparable with that measured on the best samples made with HP/HT hBN encapsulation.[40] At lower energy, several narrow peaks are also clearly visible in figure 9. We emphasize that these PL lines are not due to recombinations involving defects but correspond to the radiative recombination of different exciton complexes as already identified in WSe$_2$ monolayers encapsulated with HP/HT hBN.[41–43] The peaks lying 28 and 35 meV below $X_0$ correspond to the recombination of the triplet ($X_T$) and singlet ($X_S$) charged excitons respectively.[44] The dark neutral exciton ($X_D$) lies 40 meV below $X_0$[40] whereas the charged biexciton XX$^-$ and the dark charged excitons X$_D^-$ recombine with an energy 48 meV and 57 meV below X$_0$ respectively.[45] Finally the T$_1$ peak, lying 72 meV below X$_0$, has not yet been identified but it is also observed in WSe$_2$ monolayers encapsulated with HP/HT boron nitride.[41] The clear observation of all these exciton complexes in PL spectroscopy is again a proof of the excellent quality of the samples obtained with BNNSs encapsulations. Similar to the results for MoSe$_2$ monolayers, the detection of charged exciton complexes demonstrates that the monolayer is slightly doped. The observation of the two trions proves that this doping is of N type (because of the specific band structure of the TMDs, only one type of charged exciton can be observed in the case of a P type doping). In comparison with the PL spectra measured in charge adjustable WSe$_2$ monolayers,[41] we can infer that the doping density is typically of the order of $10^{11}$ cm$^{-2}$. Finally, we have carefully checked that the laser excitation of the BNNSs only (without any TMDs) does not yield any luminescence due to defect recombination in the visible region of the spectrum. This is an additional proof of the high quality of the BNNSs obtained with the polymer route, which confirms other positive results recently obtained on Metal-hBN-Metal capacitor devices.[46]

**Conclusion**

In this work, it has been demonstrated that the reported disruptive PDC/PCS method coupling the PDCs route with a pressure-controlled sintering has been efficiently employed for the synthesis of millimeter-scale hBN crystals. The high crystalline and chemical quality of these crystals is attested by different complementary characterization techniques including XRD, Raman spectroscopy, CL and XPS. All these methods lead to the same conclusion, namely that the quality of resulting crystals is comparable to that obtained by the HP/HT or AP/HT growth processes, with the advantage that the current process is softer, less energy consuming and scalable. Importantly, the good characteristics of these crystals are preserved after exfoliation, opening the way to the use of the extracted BNNSs for fabricating diverse devices. As a proof of concept, we showed a first successful application with optoelectronic devices made of MoSe$_2$ and WSe$_2$ monolayers encapsulated into BNNSs. The photoluminescence of TMDC layers evidences exciton complexes proving the excellent quality of the encapsulated material.

**Acknowledgements**

The authors thank the CLYM, for providing access to the SEM facilities, CECOMO for access to Raman spectroscopy and Science et Surface (Ecully, France) for XPS analyses.

This work has been partially financially supported by the European Union Horizon 2020 Program under the Graphene Flagship (Graphene Core 3, grant number 881603), the iMUST LABEX program MUSCAT-2D and the National Research Agency, France (project n° ANR-16-CE08-0021-01). Y. Li acknowledges the China Scholarship Council (CSC) for the PhD grant support.

**References**

(1) Vuong, T. Q. P.; Cassabois, G.; Valvin, P.; Rousseau, E.; Summerfield, A.; Mellor, C. J.; Cho, Y.; Cheng, T. S.; Albar, J. D.; Eaves, L.; Foxon, C. T.; Beton, P. H.; Novikov, S. V.; Gil, B. Deep Ultraviolet Emission in Hexagonal Boron Nitride Grown by High-Temperature Molecular Beam Epitaxy. *2D Mater.* **2017**, *4* (2), 021023. https://doi.org/10.1088/2053-1583/aa604a.
(2) Jiang, H. X.; Lin, J. Y. Hexagonal Boron Nitride for Deep Ultraviolet Photonic Devices. *Semicond. Sci. Technol.* **2014**, *29* (8), 084003. https://doi.org/10.1088/0268-1242/29/8/084003.
(3) Rigosi, A. F.; Levy, A. L.; Snure, M. R.; Glavin, N. R. Turn of the Decade: Versatility of 2D Hexagonal Boron Nitride. *J. Phys. Mater.* **2021**, *4* (3), 032003. https://doi.org/10.1088/2515-7639/abf1ab.
(4) Cadiz, F.; Courtade, E.; Robert, C.; Wang, G.; Shen, Y.; Cai, H.; Taniguchi, T.; Watanabe, K.; Carrere, H.; Lagarde, D.; Manca, M.; Amand, T.; Renucci, P.; Tongay, S.; Marie, X.; Urbaszek, B. Excitonic Linewidth Approaching the Homogeneous Limit in MoS 2 -Based van Der Waals Heterostructures. *Phys. Rev. X* **2017**, *7* (2), 021026. https://doi.org/10.1103/PhysRevX.7.021026.
(5) Ajayi, O. A.; Ardelean, J. V.; Shepard, G. D.; Wang, J.; Antony, A.; Taniguchi, T.; Watanabe, K.; Heinz, T. F.; Strauf, S.; Zhu, X.-Y.; Hone, J. C. Approaching the Intrinsic Photoluminescence Linewidth in Transition Metal Dichalcogenide Monolayers. *2D Mater.* **2017**, *4* (3), 031011. https://doi.org/10.1088/2053-1583/aa6aa1.
(6) Movva, H. C. P.; Fallahazad, B.; Kim, K.; Larentis, S.; Taniguchi, T.; Watanabe, K.; Banerjee, S. K.; Tutuc, E. Density-Dependent Quantum Hall States and Zeeman Splitting in Monolayer and Bilayer WSe 2. *Phys. Rev. Lett.* **2017**, *118* (24), 247701. https://doi.org/10.1103/PhysRevLett.118.247701.
(7) Pisoni, R.; Kormányos, A.; Brooks, M.; Lei, Z.; Back, P.; Eich, M.; Overweg, H.; Lee, Y.; Rickhaus, P.; Watanabe, K.; Taniguchi, T.; Imamoglu, A.; Burkard, G.; Ihn, T.; Ensslin, K. Interactions and Magnetotransport through Spin-Valley Coupled Landau Levels in Monolayer MoS 2. *Phys. Rev. Lett.* **2018**, *121* (24), 247701. https://doi.org/10.1103/PhysRevLett.121.247701.
(8) Raja, A.; Waldecker, L.; Zipfel, J.; Cho, Y.; Brem, S.; Ziegler, J. D.; Kulig, M.; Taniguchi, T.; Watanabe, K.; Malic, E.; Heinz, T. F.; Berkelbach, T. C.; Chernikov, A. Dielectric Disorder in Two-Dimensional Materials. *Nat. Nanotechnol.* **2019**, *14* (9), 832–837. https://doi.org/10.1038/s41565-019-0520-0.
(9) Taniguchi, T.; Watanabe, K. Synthesis of High-Purity Boron Nitride Single Crystals under High Pressure by Using Ba–BN Solvent. *J. Cryst. Growth* **2007**, *303* (2), 525–529. https://doi.org/10.1016/j.jcrysgro.2006.12.061.
(10) Taniguchi, T.; Watanabe, K. Synthesis of High-Purity Boron Nitride Single Crystals under High Pressure by Using Ba–BN Solvent. *J. Cryst. Growth* **2007**, *303* (2), 525–529. https://doi.org/10.1016/j.jcrysgro.2006.12.061.
(11) Watanabe, K.; Taniguchi, T.; Kanda, H. Direct-Bandgap Properties and Evidence for Ultraviolet Lasing of Hexagonal Boron Nitride Single Crystal. *Nat. Mater.* **2004**, *3* (6), 404–409. https://doi.org/10.1038/nmat1134.

(12) Zastrow, M. Meet the Crystal Growers Who Sparked a Revolution in Graphene Electronics. *Nature* **2019**, *572* (7770), 429–432. https://doi.org/10.1038/d41586-019-02472-0.
(13) Maestre, C.; Toury, B.; Steyer, P.; Garnier, V.; Journet, C. Hexagonal Boron Nitride: A Review on Selfstanding Crystals Synthesis towards 2D Nanosheets. *J. Phys. Mater.* **2021**, *4* (4), 044018. https://doi.org/10.1088/2515-7639/ac2b87.
(14) Li, J.; Elias, C.; Ye, G.; Evans, D.; Liu, S.; He, R.; Cassabois, G.; Gil, B.; Valvin, P.; Liu, B.; Edgar, J. H. Single Crystal Growth of Monoisotopic Hexagonal Boron Nitride from a Fe–Cr Flux. *J. Mater. Chem. C* **2020**, *8* (29), 9931–9935. https://doi.org/10.1039/D0TC02143A.
(15) Hoffman, T. B.; Clubine, B.; Zhang, Y.; Snow, K.; Edgar, J. H. Optimization of Ni–Cr Flux Growth for Hexagonal Boron Nitride Single Crystals. *J. Cryst. Growth* **2014**, *393*, 114–118. https://doi.org/10.1016/j.jcrysgro.2013.09.030.
(16) Liu, S.; He, R.; Ye, Z.; Du, X.; Lin, J.; Jiang, H.; Liu, B.; Edgar, J. H. Large-Scale Growth of High-Quality Hexagonal Boron Nitride Crystals at Atmospheric Pressure from an Fe–Cr Flux. *Cryst. Growth Des.* **2017**, *17* (9), 4932–4935. https://doi.org/10.1021/acs.cgd.7b00871.
(17) Yuan, S.; Linas, S.; Journet, C.; Steyer, P.; Garnier, V.; Bonnefont, G.; Brioude, A.; Toury, B. Pure & Crystallized 2D Boron Nitride Sheets Synthesized via a Novel Process Coupling Both PDCs and SPS Methods. *Sci. Rep.* **2016**, *6*, 20388. https://doi.org/10.1038/srep20388.
(18) Li, Y.; Garnier, V.; Journet, C.; Barjon, J.; Loiseau, A.; Stenger, I.; Plaud, A.; Toury, B.; Steyer, P. Advanced Synthesis of Highly Crystallized Hexagonal Boron Nitride by Coupling Polymer-Derived Ceramics and Spark Plasma Sintering Processes-Influence of the Crystallization Promoter and Sintering Temperature. *Nanotechnology* **2019**, *30* (3). https://doi.org/10.1088/1361-6528/aaebb4.
(19) Matsoso, B.; Hao, W.; Li, Y.; Vuillet-a-Ciles, V.; Garnier, V.; Steyer, P.; Toury, B.; Marichy, C.; Journet, C. Synthesis of Hexagonal Boron Nitride 2D Layers Using Polymer Derived Ceramics Route and Derivatives. *J. Phys. Mater.* **2020**, *3* (3), 034002. https://doi.org/10.1088/2515-7639/ab854a.
(20) Li, Y.; Garnier, V.; Steyer, P.; Journet, C.; Toury, B. Millimeter-Scale Hexagonal Boron Nitride Single Crystals for Nanosheet Generation. *ACS Appl. Nano Mater.* **2020**, *3* (2), 1508–1515. https://doi.org/10.1021/acsanm.9b02315.
(21) Yuan, S.; Toury, B.; Journet, C.; Brioude, A. Synthesis of Hexagonal Boron Nitride Graphene-like Few Layers. *Nanoscale* **2014**, *6* (14), 7838–7841. https://doi.org/10.1039/C4NR01017E.
(22) Castellanos-Gomez, A.; Buscema, M.; Molenaar, R.; Singh, V.; Janssen, L.; van der Zant, H. S. J.; Steele, G. A. Deterministic Transfer of Two-Dimensional Materials by All-Dry Viscoelastic Stamping. *2D Mater.* **2014**, *1* (1), 011002. https://doi.org/10.1088/2053-1583/1/1/011002.
(23) Schué, L.; Sponza, L.; Plaud, A.; Bensalah, H.; Watanabe, K.; Taniguchi, T.; Ducastelle, F.; Loiseau, A.; Barjon, J. Bright Luminescence from Indirect and Strongly Bound Excitons in H-BN. *Phys. Rev. Lett.* **2019**, *122* (6), 067401. https://doi.org/10.1103/PhysRevLett.122.067401.
(24) Schué, L.; Berini, B.; Betz, A. C.; Plaçais, B.; Ducastelle, F.; Barjon, J.; Loiseau, A. Dimensionality Effects on the Luminescence Properties of HBN. *Nanoscale* **2016**, *8* (13), 6986–6993. https://doi.org/10.1039/C6NR01253A.
(25) Cullity, B. D.; Stock, S. R. *Elements of X-Ray Diffraction*, 3. ed.; Prentice Hall: Upper Saddle River, NJ, 2001.
(26) Kuzuba, T.; Era, K.; Ishii, T.; Sato, T. A Low Frequency Raman-Active Vibration of Hexagonal Boron Nitride. *Solid State Commun.* **1978**, *25* (11), 863–865. https://doi.org/10.1016/0038-1098(78)90288-0.
(27) Pakdel, A.; Bando, Y.; Golberg, D. Nano Boron Nitride Flatland. *Chem. Soc. Rev.* **2014**, *43* (3), 934–959. https://doi.org/10.1039/C3CS60260E.
(28) Stenger, I.; Schué, L.; Boukhicha, M.; Berini, B.; Plaçais, B.; Loiseau, A.; Barjon, J. Low Frequency Raman Spectroscopy of Few-Atomic-Layer Thick HBN Crystals. *2D Mater.* **2017**, *4* (3), 031003. https://doi.org/10.1088/2053-1583/aa77d4.
(29) Gorbachev, R. V.; Riaz, I.; Nair, R. R.; Jalil, R.; Britnell, L.; Belle, B. D.; Hill, E. W.; Novoselov, K. S.; Watanabe, K.; Taniguchi, T.; Geim, A. K.; Blake, P. Hunting for Monolayer Boron Nitride: Optical and Raman Signatures. *Small* **2011**, *7* (4), 465–468. https://doi.org/10.1002/smll.201001628.

(30) Nemanich, R. J.; Solin, S. A.; Martin, R. M. Light Scattering Study of Boron Nitride Microcrystals. *Phys. Rev. B* **1981**, *23* (12), 6348–6356. https://doi.org/10.1103/PhysRevB.23.6348.

(31) Schué, L.; Stenger, I.; Fossard, F.; Loiseau, A.; Barjon, J. Characterization Methods Dedicated to Nanometer-Thick HBN Layers. *2D Mater.* **2017**, *4* (1), 015028. https://doi.org/10.1088/2053-1583/4/1/015028.

(32) Roux, S.; Arnold, C.; Paleari, F.; Sponza, L.; Janzen, E.; Edgar, J. H.; Toury, B.; Journet, C.; Garnier, V.; Steyer, P.; Taniguchi, T.; Watanabe, K.; Ducastelle, F.; Loiseau, A.; Barjon, J. Radiative Lifetime of Free Excitons in Hexagonal Boron Nitride. *Phys. Rev. B* **2021**, *104* (16), L161203. https://doi.org/10.1103/PhysRevB.104.L161203.

(33) Liu, S.; He, R.; Xue, L.; Li, J.; Liu, B.; Edgar, J. H. Single Crystal Growth of Millimeter-Sized Monoisotopic Hexagonal Boron Nitride. *Chem. Mater.* **2018**, *30* (18), 6222–6225. https://doi.org/10.1021/acs.chemmater.8b02589.

(34) Kubota, Y.; Watanabe, K.; Tsuda, O.; Taniguchi, T. Deep Ultraviolet Light-Emitting Hexagonal Boron Nitride Synthesized at Atmospheric Pressure. *Science* **2007**, *317* (5840), 932–934. https://doi.org/10.1126/science.1144216.

(35) Berns, D. H.; Cappelli, M. A. Cubic Boron Nitride Synthesis in Low-density Supersonic Plasma Flows. *Appl. Phys. Lett.* **1996**, *68* (19), 2711–2713. https://doi.org/10.1063/1.115573.

(36) Trehan, R.; Lifshitz, Y.; Rabalais, J. W. Auger and X-ray Electron Spectroscopy Studies of HBN, CBN, and N+2 Ion Irradiation of Boron and Boron Nitride. *J. Vac. Sci. Technol. Vac. Surf. Films* **1998**, *8* (6), 4026. https://doi.org/10.1116/1.576471.

(37) Rogers, C.; Gray, D.; Bogdanowicz, N.; Taniguchi, T.; Watanabe, K.; Mabuchi, H. Coherent Feedback Control of Two-Dimensional Excitons. *Phys. Rev. Res.* **2020**, *2* (1), 012029. https://doi.org/10.1103/PhysRevResearch.2.012029.

(38) Zhou, Y.; Scuri, G.; Sung, J.; Gelly, R. J.; Wild, D. S.; De Greve, K.; Joe, A. Y.; Taniguchi, T.; Watanabe, K.; Kim, P.; Lukin, M. D.; Park, H. Controlling Excitons in an Atomically Thin Membrane with a Mirror. *Phys. Rev. Lett.* **2020**, *124* (2), 027401. https://doi.org/10.1103/PhysRevLett.124.027401.

(39) Wang, G.; Gerber, I. C.; Bouet, L.; Lagarde, D.; Balocchi, A.; Vidal, M.; Amand, T.; Marie, X.; Urbaszek, B. Exciton States in Monolayer $MoSe_2$: Impact on Interband Transitions. *2D Mater.* **2015**, *2* (4), 045005. https://doi.org/10.1088/2053-1583/2/4/045005.

(40) Wang, G.; Robert, C.; Glazov, M. M.; Cadiz, F.; Courtade, E.; Amand, T.; Lagarde, D.; Taniguchi, T.; Watanabe, K.; Urbaszek, B.; Marie, X. In-Plane Propagation of Light in Transition Metal Dichalcogenide Monolayers: Optical Selection Rules. *Phys. Rev. Lett.* **2017**, *119* (4), 047401. https://doi.org/10.1103/PhysRevLett.119.047401.

(41) He, M.; Rivera, P.; Van Tuan, D.; Wilson, N. P.; Yang, M.; Taniguchi, T.; Watanabe, K.; Yan, J.; Mandrus, D. G.; Yu, H.; Dery, H.; Yao, W.; Xu, X. Valley Phonons and Exciton Complexes in a Monolayer Semiconductor. *Nat. Commun.* **2020**, *11* (1), 618. https://doi.org/10.1038/s41467-020-14472-0.

(42) Robert, C.; Dery, H.; Ren, L.; Van Tuan, D.; Courtade, E.; Yang, M.; Urbaszek, B.; Lagarde, D.; Watanabe, K.; Taniguchi, T.; Amand, T.; Marie, X. Measurement of Conduction and Valence Bands g-Factors in a Transition Metal Dichalcogenide Monolayer. *Phys. Rev. Lett.* **2021**, *126* (6), 067403. https://doi.org/10.1103/PhysRevLett.126.067403.

(43) Liu, E.; van Baren, J.; Liang, C.-T.; Taniguchi, T.; Watanabe, K.; Gabor, N. M.; Chang, Y.-C.; Lui, C. H. Multipath Optical Recombination of Intervalley Dark Excitons and Trions in Monolayer $WSe_2$. *Phys. Rev. Lett.* **2020**, *124* (19), 196802. https://doi.org/10.1103/PhysRevLett.124.196802.

(44) Jones, A. M.; Yu, H.; Schaibley, J. R.; Yan, J.; Mandrus, D. G.; Taniguchi, T.; Watanabe, K.; Dery, H.; Yao, W.; Xu, X. Excitonic Luminescence Upconversion in a Two-Dimensional Semiconductor. *Nat. Phys.* **2016**, *12* (4), 323–327. https://doi.org/10.1038/nphys3604.

(45) Liu, E.; van Baren, J.; Lu, Z.; Altaiary, M. M.; Taniguchi, T.; Watanabe, K.; Smirnov, D.; Lui, C. H. Gate Tunable Dark Trions in Monolayer $WSe_2$. *Phys. Rev. Lett.* **2019**, *123* (2), 027401. https://doi.org/10.1103/PhysRevLett.123.027401.

(46) Pierret, A.; Mele, D.; Graef, H.; Palomo, J.; Taniguchi, T.; Watanabe, K.; Li, Y.; Toury, B.; Journet, C.; Steyer, P.; Garnier, V.; Loiseau, A.; Berroir, J.-M.; Bocquillon, E.; Fève, G.; Voisin, C.; Baudin, E.; Rosticher, M.; Plaçais, B. Dielectric Permittivity, Conductivity and Breakdown Field of Hexagonal Boron Nitride. *ArXiv220105826 Cond-Mat* **2022**.